# Magnetic-field-tunable commensurate multi-$q$ charge orders on UTe$_2$ (011) surface

Yuanji Li[1,#], Ruotong Yin[1,#], Jiashuo Gong[1#], Dengpeng Yuan[2,#], Yuguang Wang[1], Shiyuan Wang[1], Mingzhe Li[1], Jiakang Zhang[1], Ziwei Xue[2], Zengyi Du[3], Shiyong Tan[2,*], Dong-Lai Feng[3,*], Ya-Jun Yan[1,3,*]

[1]*Hefei National Research Center for Physical Sciences at the Microscale and Department of Physics, University of Science and Technology of China, Hefei, 230026, China*
[2]*National Key Laboratory of Surface Physics and Chemistry, Mianyang, 621908, China*
[3]*New Cornerstone Science Laboratory, Hefei National Laboratory, Hefei, 230088, China*

[#] *Those authors contributed equally to this work.*
*E-mails: sytan4444@163.com, dlfeng@hfnl.cn, yanyj87@ustc.edu.cn;

The heavy-fermion superconductor UTe$_2$ has attracted intense interest as a candidate for spin-triplet pairing. Recent scanning tunneling microscopy (STM) studies have reported complex charge orders (COs) on its (011) surface, but their origin and relationship with superconductivity remain controversial. Here, by performing temperature-, magnetic field-, and sample-dependent STM measurements, we identify multiple new CO wave vectors beyond those previously reported. All these CO wave vectors are strictly locked to integer multiples of 1/14 and 1/4 of the reciprocal lattice vectors of the UTe$_2$ (011) surface, and multiple of them coexist in real space, collectively revealing a family of field-tunable, commensurate multi-$q$ COs. These COs exist within an energy range much larger than the superconducting energy scale, their emergence suppresses the density of states near E$_F$, yet show negligible coupling to bulk superconductivity and magnetic vortices. Our findings strongly disfavor the Fermi surface nesting or primary pair-density-wave pictures, but are consistent with a surface parent spin order.

## Introduction

A defining characteristic of strongly correlated systems is the widespread emergence of modulated orders in charge, spin, and Cooper-pair channels, giving rise to exotic phenomena such as intertwined and vestigial orders [1,2]. In superconductors, such orders sometimes take the form of pair-density waves (PDWs) [3]. Scanning tunneling microscopy (STM) has revealed PDWs in high-temperature cuprate superconductors [4,5], iron-based superconductors [6,7], NbSe$_2$ [8], Kagome superconductors [9,10], etc. PDWs can act either as a primary order that induces secondary orders such as charge density waves (CDWs), or as a secondary order, for example, resulting from the interplay between superconductivity and CDWs.

Recently, STM studies on the spin-triplet superconductor candidate UTe$_2$ reported a multi-component spin-triplet PDW, with its wave vectors identical to those of a CDW [11]. This CDW, comprising three incommensurate components denoted as $q_{1L}$, $q_{1M}$ and $q_{1R}$, exhibits unusual magnetic-field dependence and appears to be suppressed near the bulk superconducting upper critical field [12,13]. This observation suggests that superconductivity and CDW are inter-connected through a primary PDW [12]. However, subsequent STM studies revealed that the CDW persists well above the superconducting transition temperature, inconsistent with the primary PDW picture [14]. Moreover, additional CDW wave vectors denoted as $q_{3L}$, $q_{2M}$ and $q_{3R}$ were reported at ~80 mK and 10–20 T, which coexist with $q_{1L}$, $q_{1M}$ and $q_{1R}$ but are half their magnitudes [15]. These new wave vectors were interpreted as the primitive CDW wave vectors in UTe$_2$, likely originating from Fermi surface nesting [15]. Furthermore, although superconductivity in UTe$_2$ is a well-established bulk phase, no evidence for bulk CDW has been reported [16-19]. This raises two important issues: What is the precise origin of the complex CDW/charge orders (COs) observed on the UTe$_2$ (011) surface, and how to reconcile these seemingly contradictory experimental observations? To address these, we perform temperature-, magnetic field-, and sample-dependent STM measurements on high-quality UTe$_2$ crystals, discovering additional CO wave vectors and clarifying the overall organizational structure of the COs as well as their relationship with bulk superconductivity.

## Results
### A. Complex magnetic-field-tunable COs on UTe$_2$ (011) surface

The projected atomic structure of the UTe$_2$ (011) surface consists of chains of Te1, Te2, and U atoms aligned along the crystallographic *a*-axis (inset of Fig. 1a). Here, we define the direction perpendicular to the *a*-axis as the $b^*$-axis, with the intrachain spacing along the Te1 chain and the interchain spacing between adjacent Te1 chains denoted as $a$ and $b^*$, respectively. Figure 1a shows a representative d$I$/d$V$ map on the UTe$_2$ (011) surface of sample #1 at $T$ = 40 mK and zero magnetic field, revealing obvious charge modulations. The corresponding fast Fourier transform (FFT) image (Fig. 1b) exhibits Bragg peaks of the underlying lattice ($q_{Bragg}$ and $q_{b^*}$, blue squares) and six additional diffraction spots (orange circles and red triangles). All these spots coexist over the measured energy range from -40 to 80 meV and remain nondispersive (Supplementary Figs. 1 and 2), indicating the static nature of the charge ordering. The $q_{1L}$, $q_{1M}$, $q_{1R}$ and $q_{2M}$ wave vectors are consistent with previous STM reports [11-15], while $q_{2L}$ and $q_{2R}$ are newly discovered in this work.

Under perpendicular magnetic fields ($B_\perp$), the charge modulations on the UTe$_2$ (011) surface undergo pronounced changes (Figs. 1c-e), which are more clearly visualized in the FFT images (Figs. 1f-h); and more data are shown in Supplementary Fig. 3. Notably, the CO evolution depends only on the magnitude of $B_\perp$, and not on its direction. For $|B_\perp|$ < 3 T, the CO remains identical to that at zero field. At 3 T ⩽ $|B_\perp|$ < 6 T, a new wave vector $q_{3R}$ emerges alongside the initial ones (purple circles in Fig. 1f). At ~ 6.5 T, $q_{3R}$ vanishes entirely (Fig. 1g); concurrently, new modulations appear, manifested as large-period smectic stripes predominantly along the $b^*$-axis (Fig. 1d) and arc-shaped diffraction patterns $q_4$ in FFT image (magenta box in Fig. 1g). $q_4$ exists within a narrow field window and is completely suppressed by ~ 7.5 T, along with $q_{2L}$, $q_{2M}$, and $q_{2R}$; only weakened $q_{1L}$, $q_{1M}$, $q_{1R}$ remain (Figs. 1e,h). At this field, pronounced phase separation occurs: residual CO from $q_{1L}$, $q_{1M}$ and $q_{1R}$ confines to bright regions in Fig. 1e, fully suppressed elsewhere (see Supplementary Fig. 4 for details). Finally, charge modulations are virtually absent at 10 T. The newly observed $q_{3R}$ and $q_4$ wave vectors are also nondispersive (Supplementary Fig. 2).

Beyond their distinct field dependencies, these CO wave vectors also show markedly different temperature sensitivities. At $T$ = 4.2 K, only $q_{1L}$, $q_{1M}$ and $q_{1R}$ are observable, even under optimal field conditions for the other COs (Supplementary Fig. 5); at $T$ = 10 K, all are absent. This indicates that these COs are highly fragile against thermal fluctuations and stabilize only at sufficiently low temperatures. Based on these findings, we construct a schematic $T$–$|B_\perp|$ phase diagram for the COs in sample #1 (Fig. 1i). The detailed temperature- and field-dependent intensities for these CO wave vectors are shown in Supplementary Figs. 6 and 7.

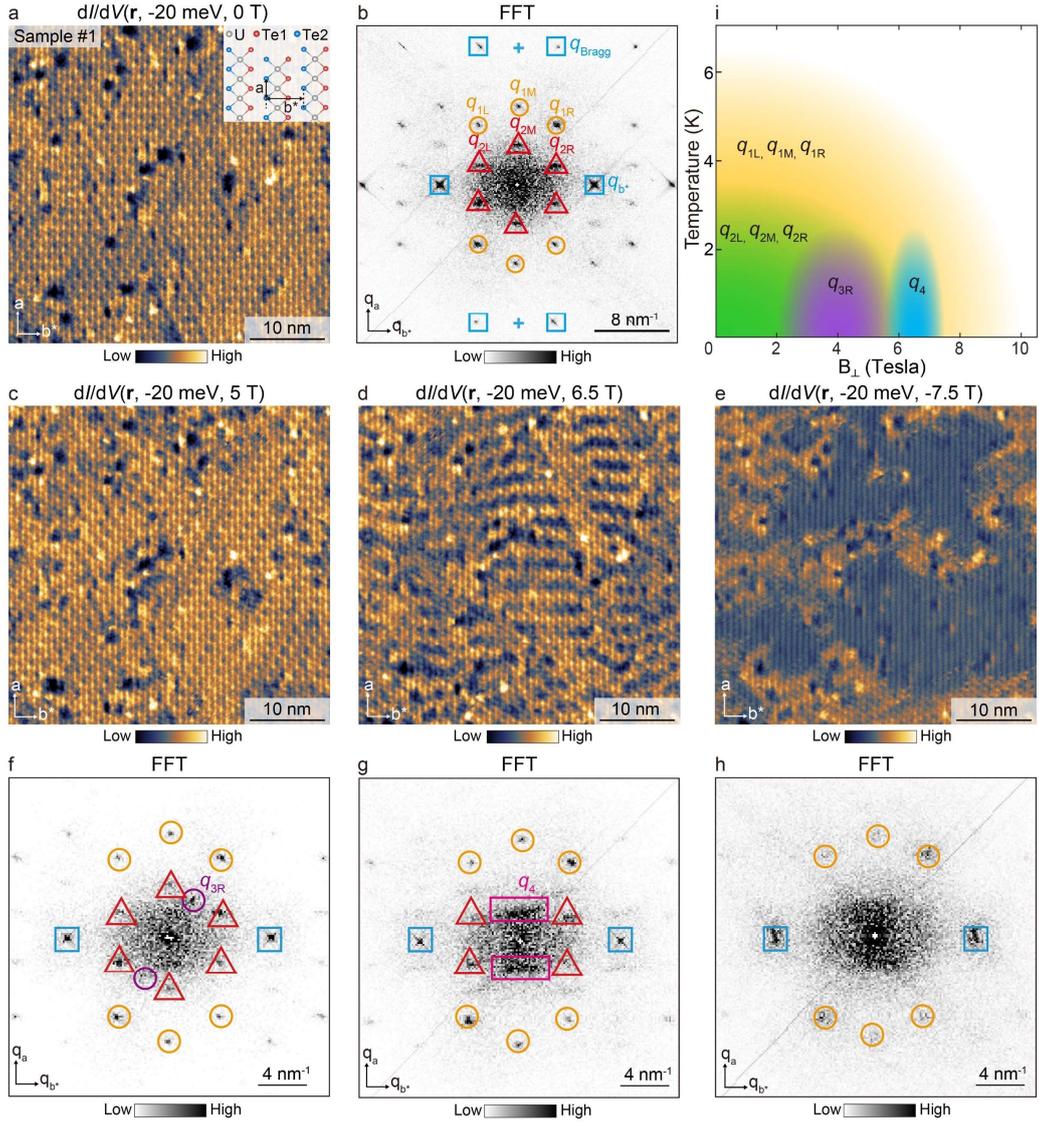

**Figure 1 | Field-tunable COs on the UTe₂ (011) surface (sample #1).** (**a**), Typical d$I$/d$V$ map acquired at $T$ = 40 mK and zero field. Inset: lattice structure of the (011) surface. (**b**), Corresponding FFT image of panel (**a**). (**c**)–(**e**), Typical d$I$/d$V$ maps in the same field of view (FOV) under $\mathbf{B}_\perp$ = 5 T, 6.5 T, and -7.5 T, respectively. (**f**)–(**h**), Corresponding FFT images of panels (**c**)–(**e**). The Bragg peaks of underlying atomic lattice, $\mathbf{q}_{Bragg}$ and $\mathbf{q}_{b^*}$, are indicated by blue rectangles, and the blue crosses indicate the positions of $\mathbf{q}_a = 2\pi/a$. The orange circles, red triangles, purple circles, and magenta rectangles mark out the positions of different CO wave vectors. (**i**), Schematic $T$–|$\mathbf{B}_\perp$| phase diagram of COs. Measurement conditions: (**a**), (**c**)–(**e**), $V_b$ = -20 mV, $I_t$ = 200 pA, $\Delta V$ = 5 mV.

### B. Comparison of COs across UTe₂ samples

To establish universality, we examine field-tunable COs across multiple UTe₂ samples. The representative data and schematic $T$–|$\mathbf{B}_\perp$| phase diagrams for samples #2 and #3 are shown in Fig. 2; additional data are shown in Supplementary Figs. 8-10.

Comparing the three samples reveals notable commonalities. At $T$ = 40 mK and zero field, all samples show the same CO wave vectors ($\mathbf{q}_{1L}$, $\mathbf{q}_{1M}$, $\mathbf{q}_{1R}$, $\mathbf{q}_{2L}$, $\mathbf{q}_{2M}$, $\mathbf{q}_{2R}$). Upon applying $\mathbf{B}_\perp$, new wave vectors emerge, such as $\mathbf{q}_{3L}$, $\mathbf{q}_{3R}$, and $\mathbf{q}_4$. The intensities of all wave vectors evolve with $\mathbf{B}_\perp$.

However, notable differences also exist in both the wave vectors involved and the critical fields for their emergence and suppression. Firstly, $q_4$ is absent in samples #2 and #3, while sample #3 hosts additional $q_{5R}$ and $q_{6R}$ wave vectors at 12 T (Figs. 2d,e). Secondly, $q_{3R}$ and its mirror-symmetric counterpart $q_{3L}$ emerge simultaneously in sample #2 at a markedly lower field threshold of 0.52 T (Figs. 2a,b) and exist in spatially segregated domains (Supplementary Fig. 9); while in samples #1 and #3, only one of them appears, likely because the measured region residing within a single large domain in these two cases. Thirdly, $q_{1L}$, $q_{1M}$, $q_{1R}$, $q_{2L}$, $q_{2M}$, $q_{2R}$, and $q_{3L}$ persist at 12 T in sample #3 (Figs. 2d,e), whereas they are fully suppressed by 10 T in sample #1.

These results suggest that the UTe$_2$ (011) surface possesses a unified ground state consisting of $q_{1L}$, $q_{1M}$, $q_{1R}$, $q_{2L}$, $q_{2M}$, and $q_{2R}$, which can be tuned by field into another state additionally comprising the energy-degenerate $q_{3L}$ or $q_{3R}$. However, variations in this tuning process across samples, along with the emergence of distinct new wave vector configurations at high fields, likely reflect differences in sample properties, such as defect types and concentrations. A detailed investigation of these variations lies beyond the scope of the present work.

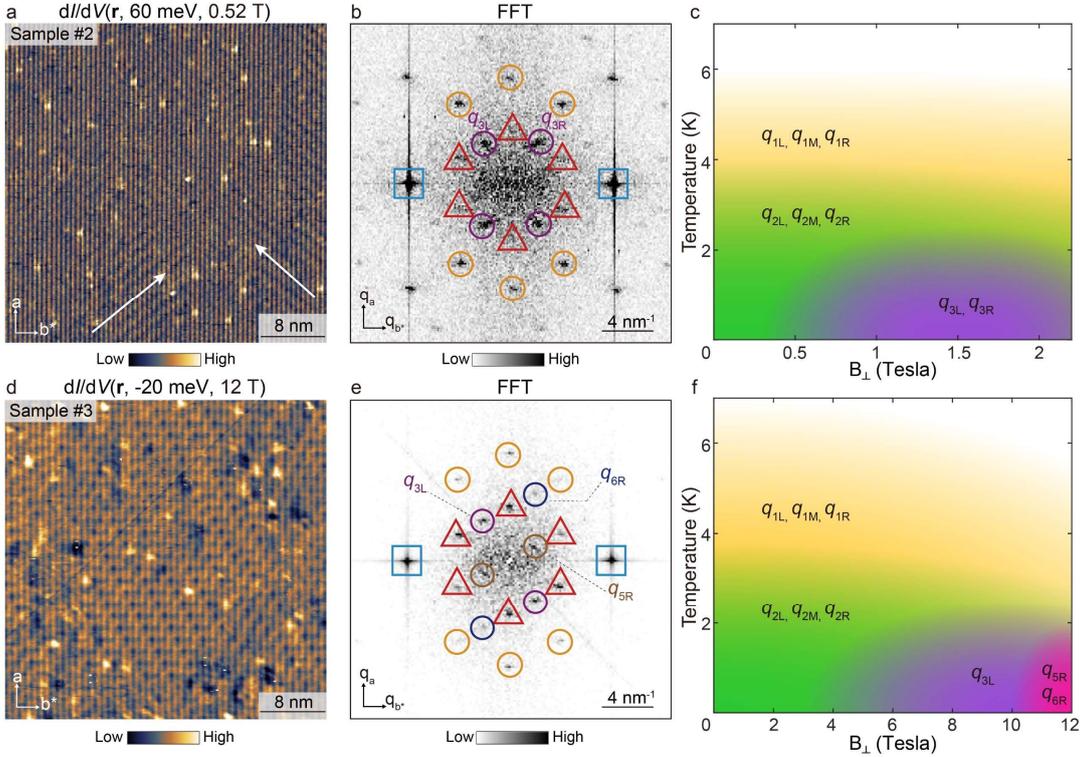

**Figure 2 | COs in UTe$_2$ samples #2 and #3.** (a),(b), Typical d$I$/d$V$ map and corresponding FFT image for sample #2 at **B**$_\perp$ = 0.52 T. The white arrows in panel (a) indicate the $q_{3R}$- and $q_{3L}$-dominated regions. (d),(e), Typical d$I$/d$V$ map and corresponding FFT image for sample #3 at **B**$_\perp$ = 12 T. Field-induced $q_{3L}$, $q_{3R}$, $q_{5R}$ and $q_{6R}$ wave vectors are marked out in panels (b),(e). (c),(f), Schematic $T$–|**B**$_\perp$| phase diagrams for COs in samples #2 and #3. Measurement conditions: (a), $V_b$ = 60 mV, $I_t$ = 80 pA, $\Delta V$ = 6 mV; (d), $V_b$ = -20 mV, $I_t$ = 200 pA, $\Delta V$ = 5 mV.

### C. Commensurability of the CO wave vectors

To further characterize these COs, we extract the magnitudes of all observed wave vectors. Figures 3b,c show FFT intensity profiles along the color-coded lines indicated in Fig. 3a. Notably, all wave vectors are locked to the underlying lattice, appearing at integer multiples of 1/14|$q_a$| and

$1/4|\mathbf{q}_{b^*}|$ along the *a*- and *b*\*-axes, respectively, where $\mathbf{q}_a = 2\pi/a$ and $\mathbf{q}_{b^*} = 2\pi/b^*$ are the reciprocal lattice vectors. For clarity, we plot all observed CO peaks on a coordinate system with $\mathbf{q}_a$ and $\mathbf{q}_{b^*}$ as basis vectors (Fig. 3d). The corresponding coordinates (in units of $(\mathbf{q}_{b^*}, \mathbf{q}_a)$) are: $\{\mathbf{q}_{1L}, \mathbf{q}_{1M}, \mathbf{q}_{1R}\}$ = $\{(-1/2, 3/7), (0, 4/7), (1/2, 3/7)\}$, $\{\mathbf{q}_{2L}, \mathbf{q}_{2M}, \mathbf{q}_{2R}\}$ = $\{(-1/2, 1/7), (0, 2/7), (1/2, 1/7)\}$, $\mathbf{q}_{3R}$ = $(1/4, 3/14)$, $\mathbf{q}_{3L}$ = $(-1/4, 3/14)$, $\mathbf{q}_4$ = $(0, 1/7)$, $\mathbf{q}_{5R}$ = $(1/4, 1/14)$, and $\mathbf{q}_{6R}$ = $(1/4, 5/14)$. Their commensurability indicates strong coupling between the COs and the atomic lattice.

The $\mathbf{q}_{1L}$, $\mathbf{q}_{1M}$ and $\mathbf{q}_{1R}$ wave vectors have the largest magnitudes and persist over the broadest range in phase diagram, whereas the magnitudes of other wave vectors are smaller and appear within distinct finite field ranges, suggesting that all wave vectors are fundamental rather than higher-order harmonics. Moreover, except for the spatially separated $\mathbf{q}_{3R}$ and $\mathbf{q}_{3L}$ domains, all wave vectors observed under the same conditions coexist in real space, indicating a multi-*q* charge modulation.

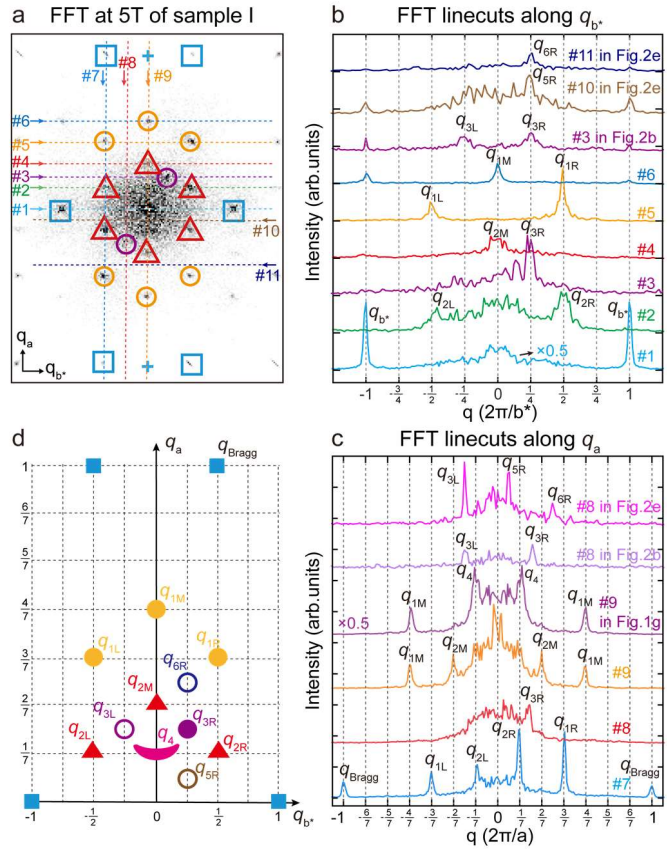

**Figure 3 | Magnitudes of the observed CO wave vectors.** (**a**), FFT image of sample #1 (the same as Fig. 1f). (**b**),(**c**), FFT intensity profiles along the color-coded lines indicated in panel (**a**). (**d**), Schematic diagram showing the distribution of multiple CO wave vectors on a coordinate system with $\mathbf{q}_a$ and $\mathbf{q}_{b^*}$ as basis vectors. Solid symbols represent CO peaks present in sample #1, while open symbols denote additional CO peaks in samples #2 and #3.

### D. Influence of CO evolution on low-energy electronic states

Having established the field-dependent COs, we next examine the response of the underlying electronic states. Figure 4a shows spatially averaged d*I*/d*V* spectra under selected fields. At zero field (bottom black curve), the density of states (DOS) drops abruptly near ±30 meV (black arrows), forming a gap-like structure, with a weak hump near -5 meV (blue arrows and inset). As $|\mathbf{B}_\perp|$ increases, the overall spectral profile remains largely unchanged, except for a gradual DOS enhancement near $E_F$ (inset of Fig. 4a). To quantify this, we integrate the DOS within ±10 meV for each field and normalize to the value at 10 T (Fig. 4b). The normalized DOS remains nearly constant for $|\mathbf{B}_\perp| \leq 5$ T, then rises for $|\mathbf{B}_\perp| > 5$ T and increases rapidly once above 7.5 T. These results indicate that the presence of COs suppresses the DOS near $E_F$, thereby lowering the electronic energy.

Under $\mathbf{B}_\perp$, elongated magnetic vortices emerge on UTe$_2$ (011) surface (Fig. 4c), characterized by a distinct dark-bright contrast in local DOS [20-22]. Comparing the vortex lattice with the spatial amplitude distribution of different CO components (Fig. 4d and Supplementary Fig. 11) reveals no discernible enhancement or suppression of CO amplitude within the vortex cores; conversely, the vortex morphology itself remains unaffected by COs. Moreover, the vortices persist even after the COs completely disappear at 10 T. Figure 4e shows representative superconducting gap spectra away from vortices; the gap is progressively suppressed with increasing $\mathbf{B}_\perp$. The gap depth, defined as $(\frac{dI}{dV}(0.25\ meV) - \frac{dI}{dV}(0\ meV))/\frac{dI}{dV}(0.25\ meV)$, decreases smoothly with field (Fig. 4f), with no abrupt changes across the phase transition fields of COs. Taken together, these observations suggest a predominantly surface-related origin of these COs, in contrast to the bulk nature of superconductivity, which may account for the weak coupling among the COs, superconductivity, and magnetic vortices in UTe$_2$.

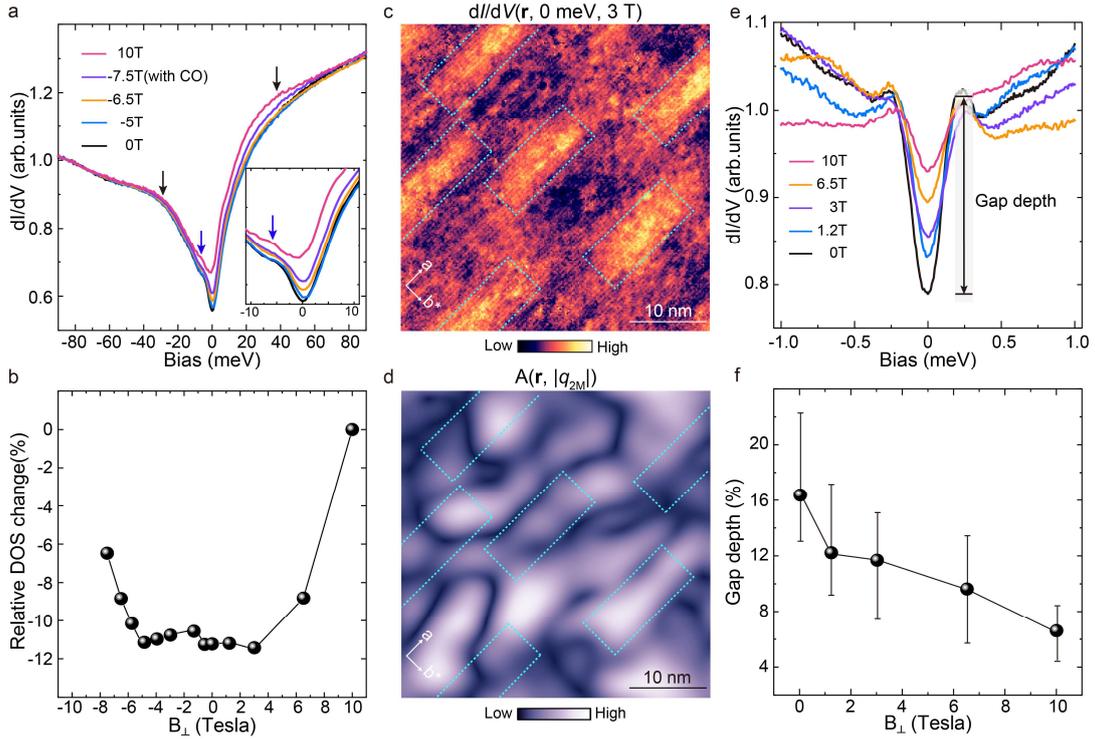

**Figure 4 | Response of low-energy electronic state and superconductivity to COs in sample #1 at $T$ = 40 mK.** (**a**), Spatially averaged d$I$/d$V$ spectra under selected fields. Inset: enlarged view near $E_F$. (**b**), Relative DOS change with field, defined as $\frac{\frac{dI}{dV}(B_\perp) - \frac{dI}{dV}(10\ T)}{\frac{dI}{dV}(10\ T)}$. The DOS is integrated within ±10 meV. (**c**),(**d**), Zero-energy magnetic vortex map and amplitude distribution of $\mathbf{q}_{2M}$ at −20 meV, acquired in the same FOV at $\mathbf{B}_\perp$=3 T. Dashed cyan rectangles mark the vortex positions. The amplitude map in panel (**d**) is obtained using a Gaussian FFT filter with width $\sigma_q \sim$ 5.5 nm. (**e**), Representative superconducting gap spectra away from vortices, which are normalized to their normal state resistances $\mathbf{R}_N = (V_b(+1\ meV) - V_b(-1\ meV))/(I_t(+1\ meV) - I_t(-1\ meV))$. (**f**), Field-dependent gap depth. Black

circles represent spatially averaged gap depth, the error bars indicate the variation range arising from spatial inhomogeneity of the superconducting gap. Measurement conditions: (**a**), $V_b$ = 90 mV, $I_t$ = 400 pA, $\Delta V$ = 2 mV; (**c**), $V_b$ = -1.5 mV, $I_t$ = 200 pA, $\Delta V$ = 0.08 mV; (**e**), $V_b$ = 1 mV, $I_t$ = 200 pA, $\Delta V$ = 0.05 mV.

## Discussion and conclusion

In this manuscript, we report a series of field-tunable CO wave vectors on the UTe$_2$ (011) surface, including the newly identified $q_{2L}$, $q_{2R}$, $q_4$, $q_{5R}$ and $q_{6R}$, and reveal their highly distinctive behaviors.

Firstly, all CO wave vectors are field-sensitive and strictly locked to integer multiples of $1/14|q_a|$ and $1/4|q_{b^*}|$ along the *a*- and *b*$^*$- axes, respectively. This precise commensurability and simple fractional relationships extend beyond conventional Fermi surface nesting picture, which typically yields one or two dominant incommensurate wave vectors. Alternative scenarios also appear unlikely. A primary PDW picture or a quasiparticle-interference origin from topological surface states[12,23,24] cannot account for the nondispersive nature of all wave vectors and their persistence over an energy range far exceeding the superconducting gap. Lattice instabilities driven by electron-phonon coupling, valence order, or orbital order typically exhibit weak field responses and are therefore implausible. Instead, the strong field dependence points toward a spin-related order, with the observed charge modulations arising as concomitant phenomena through coupling between the spin, lattice, and charge channels.

Secondly, multiple fundamental CO wave vectors typically coexist in real space, suggesting that they originate from a common parent order parameter, which, based on the preceding discussion, is most likely a spin order. Such a spin order is expected to possess a relatively complex waveform, and its intricate coupling with the lattice and charge channels gives rise to the observed multi-*q* charge modulations. At ultralow temperature and zero field, a stable spin order exists, as evidenced by the reproducible appearance of $q_{1L}$, $q_{1M}$, $q_{1R}$, $q_{2L}$, $q_{2M}$, and $q_{2R}$ wave vectors across samples. Temperature and magnetic field can significantly modify its internal spin texture, leading to changes in CO patterns and intensities. Such modulations may be understood as a fine-tuning of the underlying spin order, given that the original $q_{1L}$, $q_{1M}$, and $q_{1R}$ wave vectors persist over the widest region of the phase diagram, followed by $q_{2L}$, $q_{2M}$, and $q_{2R}$.

Thirdly, complex multi-*q* COs, or more fundamentally, the underlying spin orders that give rise to them, typically emerge in systems with intricate magnetic interactions or in proximity to quantum critical points. UTe$_2$, situated at the paramagnetic end of the uranium-based heavy-fermion family, exhibits strong magnetic anisotropy and significant spin-orbit coupling [25,26]. Although long-ranged magnetic order is absent in the bulk, complex magnetic excitations have been reported, with evidence suggesting the coexistence of ferromagnetic and antiferromagnetic fluctuations, as well as magnetic quantum critical points [27-37]. However, the COs observed in this study cannot be explained by bulk magnetic properties alone. Inelastic neutron scattering has revealed two incommensurate magnetic excitations with wave vectors of (0, 0.57, 0) and (0.41, 0.41, 0) in units of (2π/*a*, 2π/*b*, 2π/*c*), which persist up to 11 T [31,38]. These differ markedly from the field-sensitive, commensurate COs reported here. Moreover, such COs have not been observed in bulk measurements [16-19], strongly suggesting that they originate from the surface.

On the cleaved UTe$_2$ (011) surface, the local chemical environment can be modified by symmetry breaking, lattice relaxation, or chemical reconstruction. These changes can alter the U valence, crystal-field effects, correlation strength, and magnetic exchange interactions in the surface layers relative to the bulk. Previous studies have suggested a change in U valence at the surface [39,40]. Such surface effects may collectively stabilize emergent spin-related orders, finally giving rise to the observed COs in STM measurements. Consistent with this picture, we find that the emergence of the COs lowers the electronic energy of the system (Figs. 4a,b), thereby promoting their stability. Surface-induced magnetism variations are common in 4$f$/5$f$ systems [41-46]. New magnetic orders have been observed on the surfaces of EuRh$_2$Si$_2$, EuIr$_2$Si$_2$, and TbRh$_2$Si$_2$ [43-45], and the magnetic ordering temperature at the Gd(0001) surface is notably higher than in the bulk [42]. To directly test the proposed spin order, surface-sensitive magnetic probes such as spin-polarized STM and spin-polarized low-energy electron diffraction will be necessary in further studies.

In summary, our study reveals a family of field-tunable, commensurate multi-$q$ COs on the UTe$_2$ (011) surface that are surface-derived, highly reproducible, and largely decoupled from bulk superconductivity. These results provide further constraints for understanding the origin of these COs and point toward a surface spin-related parent state. The COs are sensitive to multiple dimensions, including magnetic field, temperature, and surface condition, suggesting the presence of competing orders with nearly-degenerate energies and strong fluctuations. More broadly, these complex COs, and more fundamentally the underlying spin-related orders, unravel new aspects of the magnetic interactions and electronic correlations in UTe$_2$, calling for dedicated theoretical investigations.

**Methods**

**STM measurements.** UTe$_2$ crystals were mechanically cleaved at 80 K in ultrahigh vacuum with a base pressure better than $2 \times 10^{-10}$ mbar and immediately transferred into UNISOKU-1600 STM. Pt-Ir tips were used for STM measurements after being treated on a clean Au (111) substrate. The d$I$/d$V$ spectra were collected by a standard lock-in technique with a modulation frequency of 973 Hz and a modulation amplitude $\Delta V$ of 0.05 to 6 mV. The effective electron temperature ($T_{\text{eff}}$) of our STM is ~ 170 mK when the sample temperature is ~ 40 mK [47].

**Data availability**
The main data supporting the findings of this study are available within the article and its Supplementary Information files. All the raw data generated in this study are available from the corresponding author upon request.

**Code availability**
All the data analysis codes related to this study are available from the corresponding author upon request.

**Acknowledgments**
We thank Prof. Rui Wang and Prof. Tong Zhang for helpful discussions. This work is supported by National Key R&D Program of the MOST of China (Grants No. 2023YFA1406304 (Y.J.Y.)), the National Natural Science Foundation of China (Grants No. 12374140 (Y.J.Y.), No. U23A20580

(S.Y.T.), No. 12494593 (Y.J.Y.)), the Innovation Program for Quantum Science and Technology (Grant No. 2021ZD0302803 (D.L.F.)), the New Cornerstone Science Foundation (D.L.F.), Sichuan Science and Technology Program (2025NSFJQ0040 (S.Y.T.)), Anhui Provincial Natural Science Foundation(No. 2508085J001).

**Author contributions**
UTe$_2$ single crystals were grown by D. Y., Z. X., and S. T.; STM measurements were performed by Y. L., R. Y., J. G.; Data analysis was performed by Y. L., R. Y., J. G., Y. W., S. W., J. Z., M. L., Z. D. and Y. Y.; Y. Y. and D. F. coordinated the whole work and wrote the manuscript. All authors have discussed the results and the interpretation.

**Competing interests**
The authors declare no competing interests.

# Supplementary Information for
# "Magnetic-field-tunable commensurate multi-$q$ charge orders on UTe$_2$ (011) surface"

Yuanji Li[1,#], Ruotong Yin[1,#], Jiashuo Gong[1#], Dengpeng Yuan[2,#], Yuguang Wang[1], Shiyuan Wang[1], Mingzhe Li[1], Jiakang Zhang[1], Ziwei Xue[2], Zengyi Du[3], Shiyong Tan[2,*], Dong-Lai Feng[3,*], Ya-Jun Yan[1,3,*]

[1]*Hefei National Research Center for Physical Sciences at the Microscale and Department of Physics, University of Science and Technology of China, Hefei, 230026, China*
[2]*National Key Laboratory of Surface Physics and Chemistry, Mianyang 621908, China.*
[3]*New Cornerstone Science Laboratory, Hefei National Laboratory, Hefei, 230088, China*

**Supplementary Note 1: COs on UTe$_2$ (011) surface under various energies at $B_\perp = 0$ T and $T = 40$ mK (sample #1)**

Supplementary Fig. 1 presents representative topographic image, d$I$/d$V$ maps, and the corresponding FFT images of the UTe$_2$ (011) surface in sample #1, acquired within the same field of view (FOV) at $B_\perp = 0$ T and $T = 40$ mK. While the CO modulations are weak in the topographic image (Supplementary Fig. 1a1), they are clearly visible in the d$I$/d$V$ maps (Supplementary Figs. 1b1-h1). The corresponding FFT images reveal the coexistence of $\mathbf{q}_{1L}$, $\mathbf{q}_{1M}$, $\mathbf{q}_{1R}$, $\mathbf{q}_{2L}$, $\mathbf{q}_{2M}$, $\mathbf{q}_{2R}$ wave vectors across the entire measured energy range from -40 meV to 80 meV (Supplementary Figs. 1a2-h2).

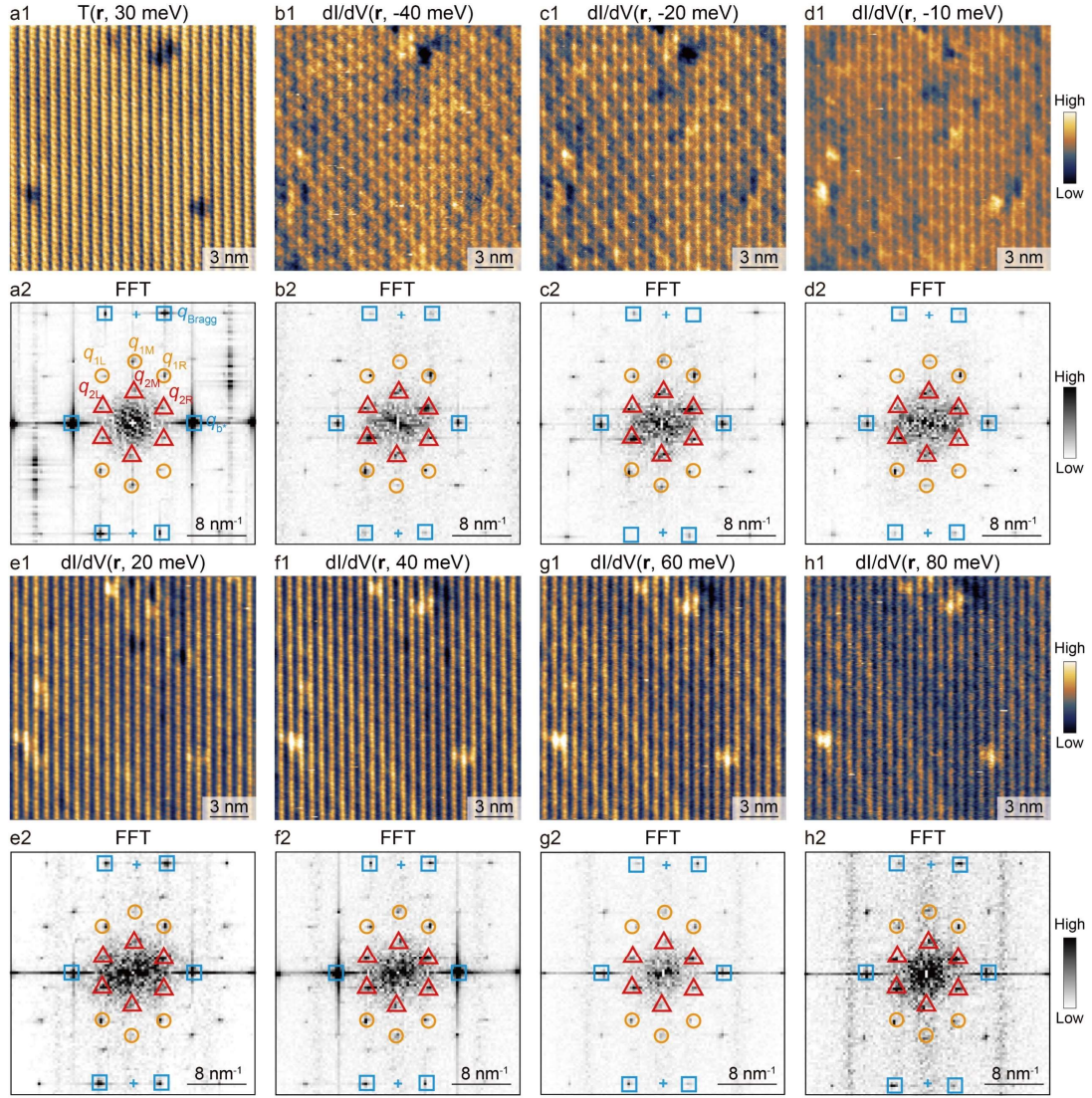

**Supplementary Figure 1 | COs on the UTe$_2$ (011) surface (sample #1) at $B_\perp = 0$ T and $T = 40$ mK.** (**a1**),(**a2**), Representative topographic image and its corresponding FFT image. (**b1**)-(**h1**), d$I$/d$V$ maps acquired at various energies within the same FOV of panel (**a1**). (**b2**)-(**h2**), Corresponding FFT images of panels (**b1**)-(**h1**). Measurement conditions: (**a1**), $V_b = 30$ mV, $I_t = 200$ pA; (**b1**)-(**h1**), $V_b = -40 - +80$ mV, $I_t = 200$ pA, $\Delta V = 5$ mV.

**Supplementary Note 2: Nondispersive behavior of the observed wave vectors in sample #1**

Supplementary Fig. 2a shows the representative FFT images of d$I$/d$V$ maps acquired at **B**$_\perp$ = 0 T, 5 T, and 6.5 T. Supplementary Figs. 2b-e present the energy dependence of the observed wave vectors, extracted along cuts #1-#4 marked in Supplementary Fig. 2a. All wave vectors are nondispersive, confirming their origin as static COs.

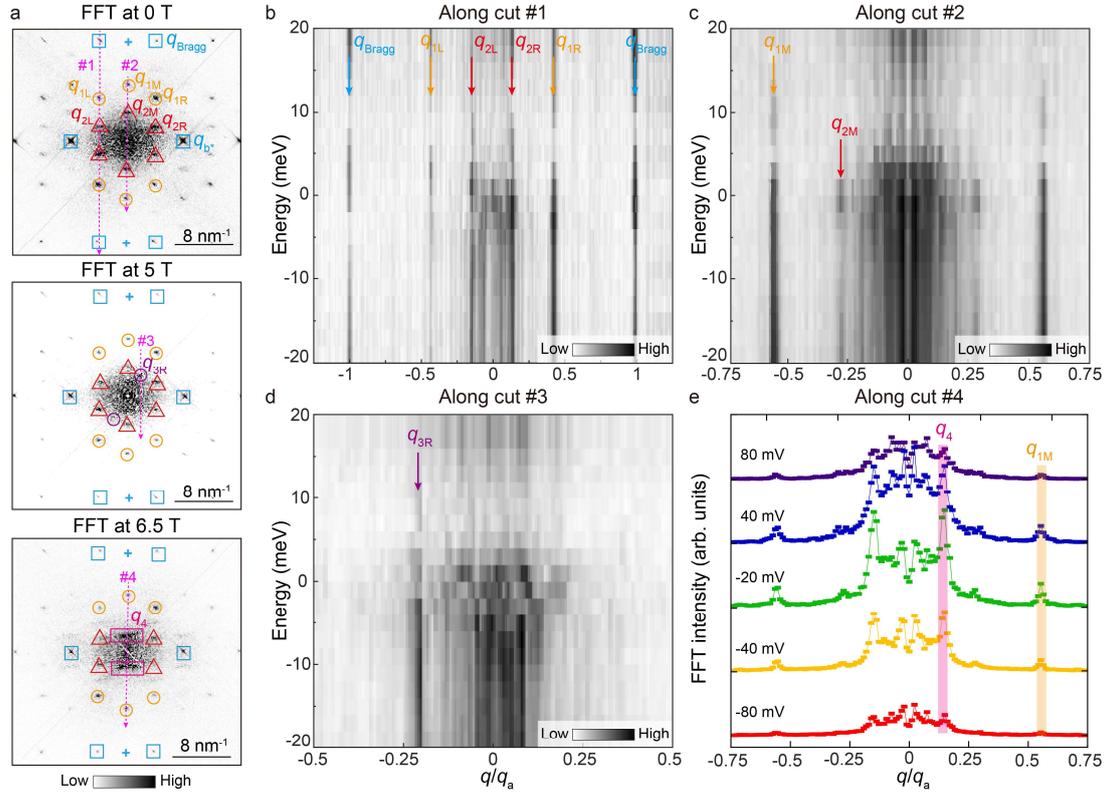

**Supplementary Figure 2 | Nondispersive behavior of the observed wave vectors in sample #1.** (**a**), Typical FFT images of the d$I$/d$V$ maps taken at B$_\perp$ = 0 T, 5 T and 6.5 T (also seen in Fig. 1 of main text). (**b**)-(**e**), Energy dependence of the wave vectors, extracted along cuts #1-#4 marked in panel (**a**).

**Supplementary Note 3: Additional datasets for B$_\perp$-dependent CO evolution in sample #1**

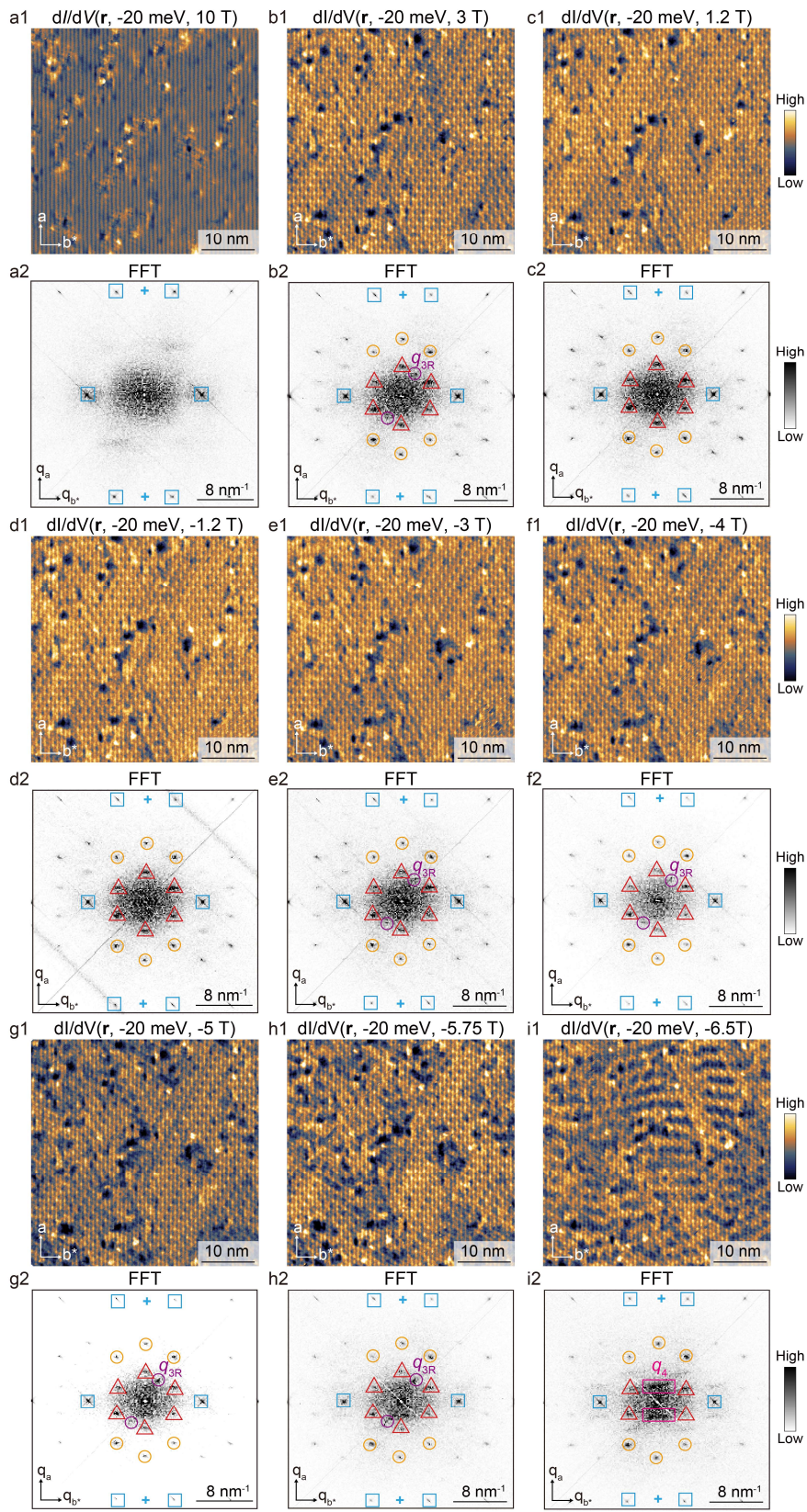

**Supplementary Figure 3 | Additional datasets for CO evolution with B$_\perp$ in sample #1, measured in the same FOV as Fig. 1a in the main text.** Measurement conditions: $V_b$ = -20 mV, $I_t$ = 200 pA, $\Delta V$ = 5 mV.

**Supplementary Note 4: Phase separation in sample #1 at $B_\perp = -7.5$ T**

In sample #1, we observe pronounced phase separation in d$I$/d$V$ maps acquired at $B_\perp = -7.5$ T and $T = 40$ mK (Supplementary Figs. 4a,b). The FOV is divided into two distinct types of domains (domains I and II) with contrasting DOS intensity; domain II, exhibiting lower DOS intensity, is outlined by white dashed curves. The corresponding FFT images (insets of Supplementary Figs. 4a,b) reveal the presence of $q_{1L}$, $q_{1M}$, and $q_{1R}$ diffraction peaks. Magnified views of domains I and II are shown in Supplementary Figs. 4c,e, respectively. Domain I exhibits obvious charge modulations, with strong $q_1$ signals in its FFT image (Supplementary Figs. 4c,d). In contrast, CO signals are virtually absent in domain II (Supplementary Figs. 4e,f). These observations reveal key insights in two respects. Firstly, they demonstrate that the $q_{1L}$, $q_{1M}$, and $q_{1R}$ wave vectors undergo phase separation at $B_\perp = -7.5$ T; Secondly, they confirm that the three wave vectors evolve in unison — appearing together with strong intensity in domain I and vanishing collectively in domain II.

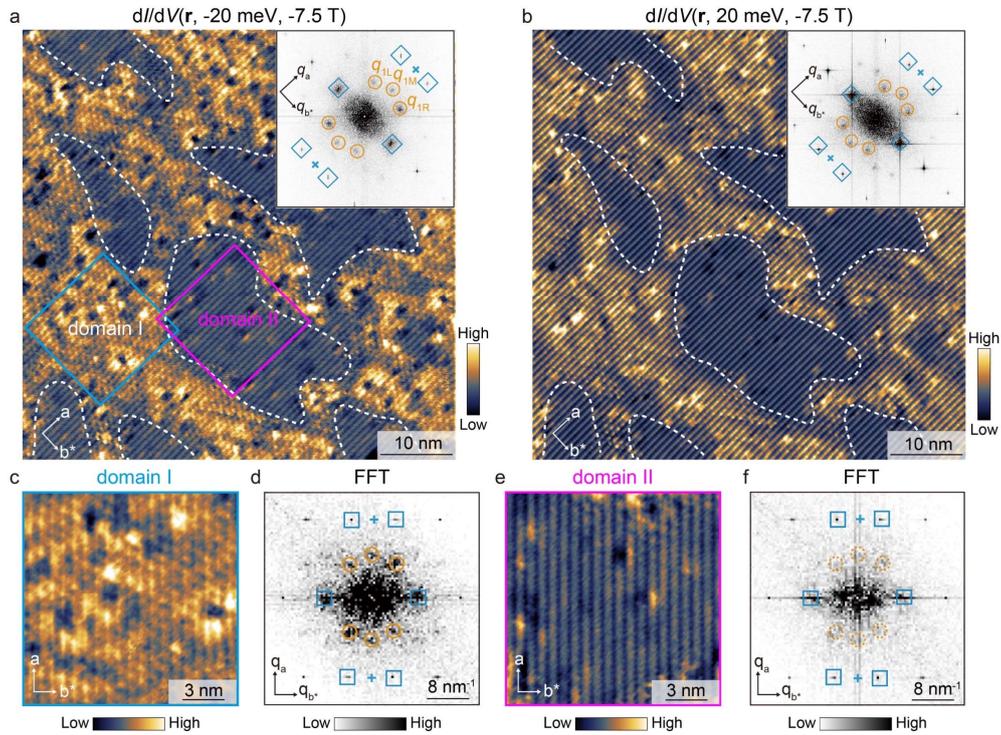

**Supplementary Figure 4 | Phase separation in sample #1 at $B_\perp = -7.5$ T. (a),(b)**, d$I$/d$V$ maps taken at E = -20 meV and +20 meV, respectively. White dashed curves outline type-II domains where CO modulations are absent. Insets: corresponding FFT images, showing only the $q_{1L}$, $q_{1M}$, and $q_{1R}$ wave vectors (orange circles). **(c)-(f)**, Magnified views of domains I and II (blue and magenta squares in panel **(a)**) and their corresponding FFT images. Measurement conditions: **(a),(b)** $V_b = -20$ mV, $I_t = 200$ pA, $\Delta V = 5$ mV.

**Supplementary Note 5: Temperature-dependent CO evolution in sample #1**

Supplementary Figs. 5a-h present representative d$I$/d$V$ maps and corresponding FFT images of sample #1, acquired under selected $B_\perp$ at $T$ = 4.2 K and 10 K. At zero field and $T$ = 4.2 K, only $q_{1L}$, $q_{1M}$, and $q_{1R}$ are discernible (Supplementary Fig. 5e), but their intensity is significantly suppressed compared to those at $T$ = 40 mK (Supplementary Figs. 5i,j). $q_{2L}$, $q_{2M}$, and $q_{2R}$ wave vectors, which coexist with $q_{1L}$, $q_{1M}$, and $q_{1R}$ at 40 mK, are almost completely suppressed (Supplementary Figs. 5e,i,j). At $T$ = 10 K, even the residual $q_{1L}$, $q_{1M}$, and $q_{1R}$ signals vanish, leaving the FFT image essentially featureless apart from the Bragg peaks (Supplementary Figs. 5b,f,i,j). As discussed earlier, applying $B_\perp$ = 5 T or 6.5 T at $T$ = 40 mK induces the $q_{3R}$ and $q_4$ wave vectors. In stark contrast, when the same fields are applied at $T$ = 4.2 K, no signatures of $q_{3R}$ or $q_4$ are observed (Supplementary Figs. 5c,d,g,h,k,l); only weak $q_1$ signals remain. The disappearance of these CO wave vectors at elevated temperatures demonstrates that they are highly fragile against thermal fluctuations and can only be stabilized at low temperatures.

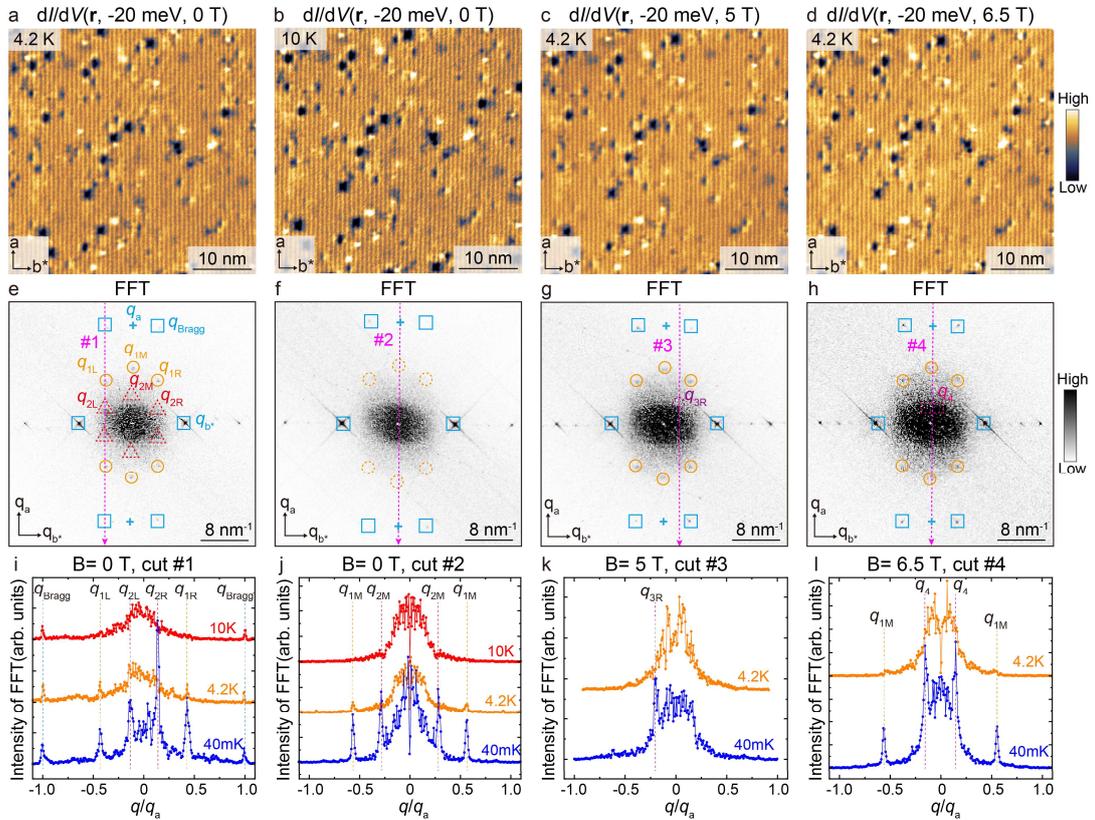

**Supplementary Figure 5 | COs at elevated temperatures for sample #1.** (**a**)-(**d**), Representative d$I$/d$V$ maps under selected magnetic fields at $T$ = 4.2 K and 10 K. (**e**)-(**h**), Corresponding FFT images of panels (**a**)-(**d**). Solid and dashed symbols indicate the presence and absence of the corresponding CO wave vectors, respectively. (**i**)-(**l**), FFT line profiles extracted along cuts #1-4 indicated in panels (**e**)-(**h**). Measurement conditions: (**a**)-(**d**), $V_b$ = -20 mV, $I_t$ = 200 pA, $\Delta V$ = 5 mV.

**Supplementary Note 6: Temperature- and field-dependent CO intensities for sample #1**

Supplementary Fig. 6 illustrates the procedure for extracting CO intensities from FFT line profiles. A representative FFT image of the d$I$/d$V$ map in sample #1 is shown in Supplementary Fig. 6a, FFT line profiles taken along cuts #1-#5 (blue arrows) are displayed in Supplementary Figs. 6b–h. For each cut, the peak intensity at the selected **q** vectors is obtained by subtracting a smooth Gaussian background from the raw profile. The resulting field- and temperature-dependent intensities of the selected CO wave vectors are presented in Supplementary Fig. 7, which serves as the basis for the schematic phase diagram in Fig. 1i of the main text.

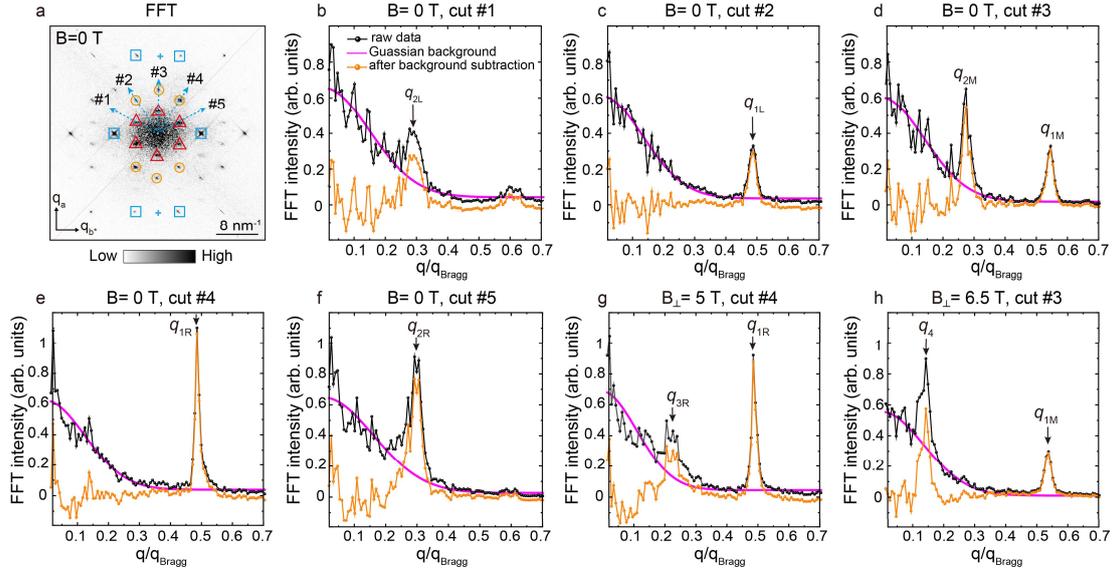

**Supplementary Figure 6 | Extraction of CO intensities from FFT line profiles.** (**a**), Representative FFT image of a d$I$/d$V$ map at zero field for sample #1 (same as Fig. 1b in the main text). Blue arrows indicate cuts #1–#5. (**b**)–(**f**), FFT line profiles extracted along cuts #1–#5 in panel (**a**). (**g**), FFT line profile along cut #4 at $B_\perp$= 5 T. (**h**), FFT line profile along cut #3 at $B_\perp$= 6.5 T. In panels (**b**)-(**h**), black curves represent the raw FFT intensity, magenta curves show the fitted Gaussian background, and orange curves display the background-subtracted intensities. The peak intensity for each labeled **q** vector is defined as the maximum of the residual signal.

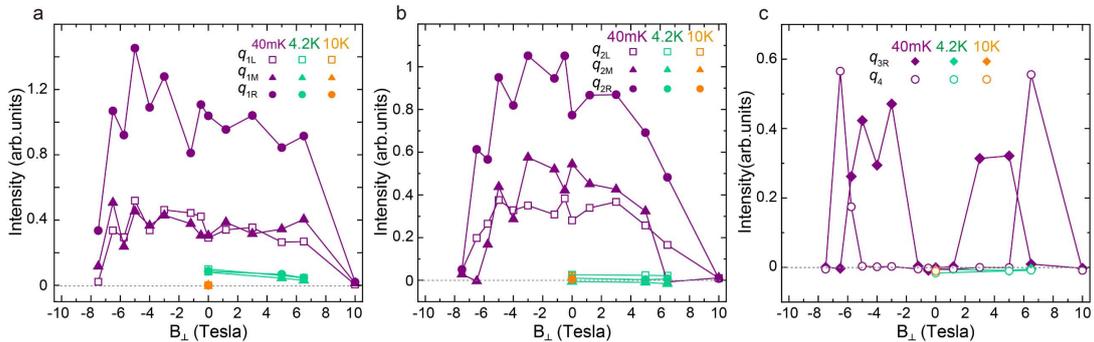

**Supplementary Figure 7 | Temperature- and magnetic-field-dependent CO intensities for sample #1.**

**Supplementary Note 7: COs across UTe$_2$ samples**

Having discussed the CO properties in sample #1, we now turn to samples #2 and #3. At $T =$ 40 mK and zero magnetic field, both sample #2 and sample #3 exhibit the same CO wave vectors as observed in sample #1, $\mathbf{q}_{1L}$, $\mathbf{q}_{1M}$, $\mathbf{q}_{1R}$, $\mathbf{q}_{2L}$, $\mathbf{q}_{2M}$, and $\mathbf{q}_{2R}$, as shown in Supplementary Figure 8.

For sample #2, upon applying $\mathbf{B}_\perp = 0.52$ T, both $\mathbf{q}_{3L}$ and $\mathbf{q}_{3R}$ emerge simultaneously (Supplementary Figs. 9a-c). To quantify their local strength, we extract the spatial amplitude distribution for each CO component[1-4], $A(\mathbf{r}, |\mathbf{q}_{3L}|)$ and $A(\mathbf{r}, |\mathbf{q}_{3R}|)$, which vary strongly on the nanoscale (Supplementary Figs. 9d,e). To determine whether $\mathbf{q}_{3L}$ and $\mathbf{q}_{3R}$ are intertwined or domain-polarized, we define a local unidirectionality parameter: $F(\mathbf{r}) = (A(\mathbf{r}, |\mathbf{q}_{3L}|) - A(\mathbf{r}, |\mathbf{q}_{3R}|))/(A(\mathbf{r}, |\mathbf{q}_{3L}|) + A(\mathbf{r}, |\mathbf{q}_{3R}|))$. Regions with $F(\mathbf{r}) > 0$ (red color) are dominated by $\mathbf{q}_{3L}$, while those with $F(\mathbf{r}) < 0$ (blue color) are dominated by $\mathbf{q}_{3R}$. As shown in Supplementary Fig. 9f, $F(\mathbf{r})$ is highly heterogeneous and forms a clear domain structure, demonstrating that $\mathbf{q}_{3L}$ and $\mathbf{q}_{3R}$ are domain-polarized rather than intertwined. This finding is consistent with other samples: only $\mathbf{q}_{3R}$ in sample #1 and only $\mathbf{q}_{3L}$ in sample #3 within the measured regions, likely due to each region residing within a single large domain.

For sample #3, $\mathbf{q}_{1L}$, $\mathbf{q}_{1M}$, $\mathbf{q}_{1R}$, $\mathbf{q}_{2L}$, $\mathbf{q}_{2M}$, and $\mathbf{q}_{2R}$ maintain at low fields (Supplementary Figs. 10a1–c1). As $\mathbf{B}_\perp$ increases to 5-7.6 T, $\mathbf{q}_{3L}$ emerges while $\mathbf{q}_{3R}$ remains absent (Supplementary Figs. 10a2–c4), consistent with the domain-polarized scenario. At $\mathbf{B}_\perp = 12$ T (Figs. 2d,e in the main text), all these wave vectors persist, accompanied by the emergence of two additional wave vectors, $\mathbf{q}_{5R}$ and $\mathbf{q}_{6R}$.

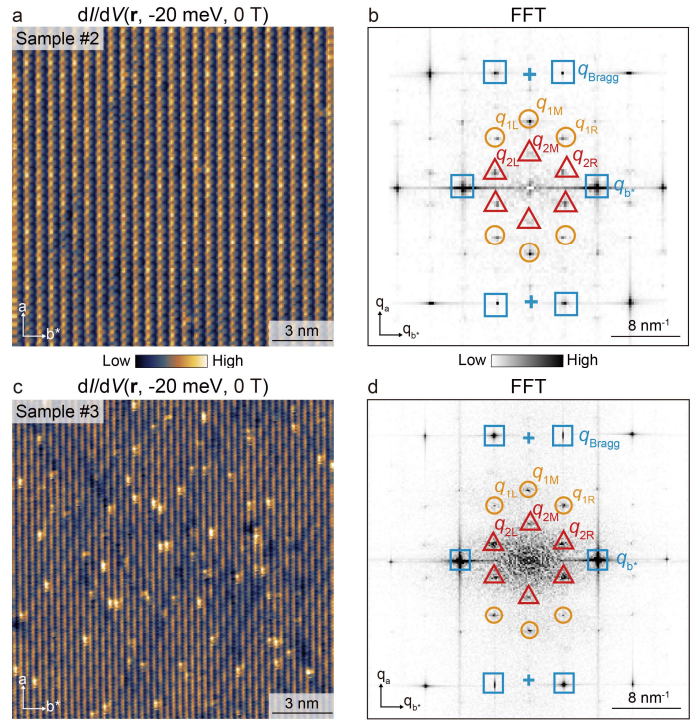

**Supplementary Figure 8 | COs in samples #2 and #3 at $T$ = 40 mK and zero magnetic field. (a)-(d),** Typical d$I$/d$V$ maps and the corresponding FFT images. Measurement conditions: **(a)**, $V_b$ = -20 mV, $I_t$ = 100 pA, $\Delta V$ = 5 mV; **(c)**, $V_b$ = -20 mV, $I_t$ = 200 pA, $\Delta V$ = 5 mV.

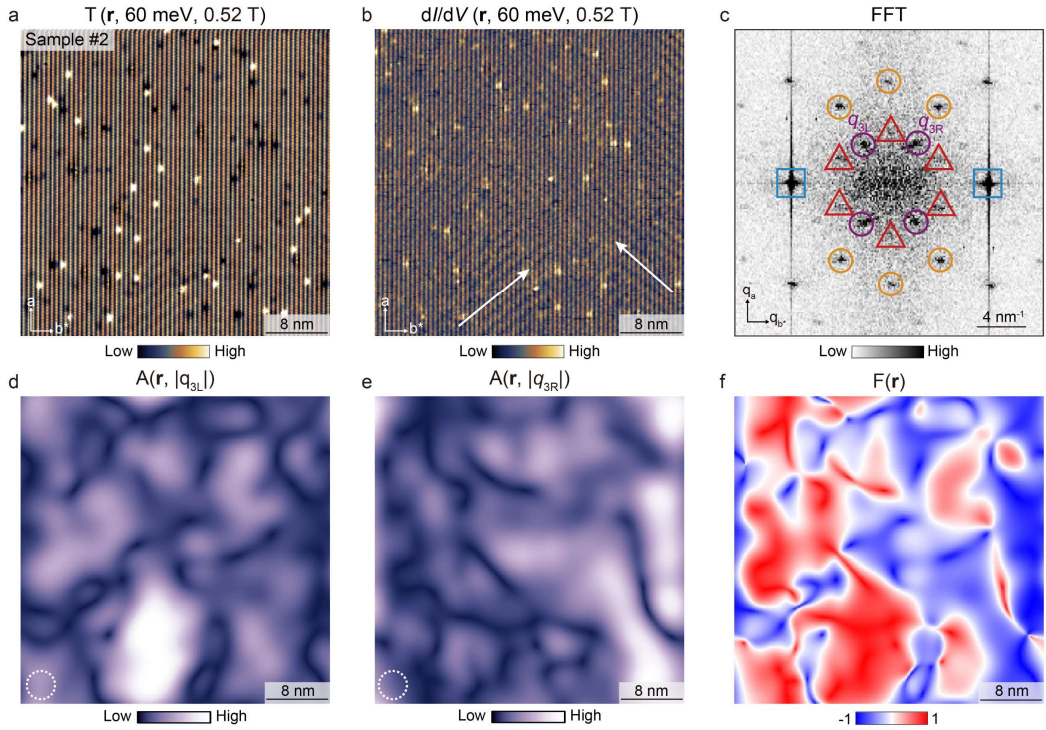

**Supplementary Figure 9 | Domain-polarized $q_{3L}$ and $q_{3R}$ wave vectors in sample #2.** (**a**), Representative topographic image of sample #2, acquired at $B_\perp = 0.52$ T and $T = 40$ mK. (**b**), Representative d$I$/d$V$ map within the same FOV as panel (**a**). (**c**), Corresponding FFT image of panel (**b**). Panels (**b**,**c**) are also shown in Fig. 2a,b of main text. (**d**),(**e**), Spatial amplitude distribution for $q_{3L}$ and $q_{3R}$ components. Dashed circles indicate the Gaussian filter width ($\sigma_q$) used to obtain the amplitude maps. (**f**), Spatial distribution of local unidirectionality parameter F(**r**). Measurement conditions: (**a**) $V_b = 60$ mV, $I_t = 100$ pA; (**b**) $V_b = 60$ mV, $I_t = 80$ pA, $\Delta V = 6$ mV.

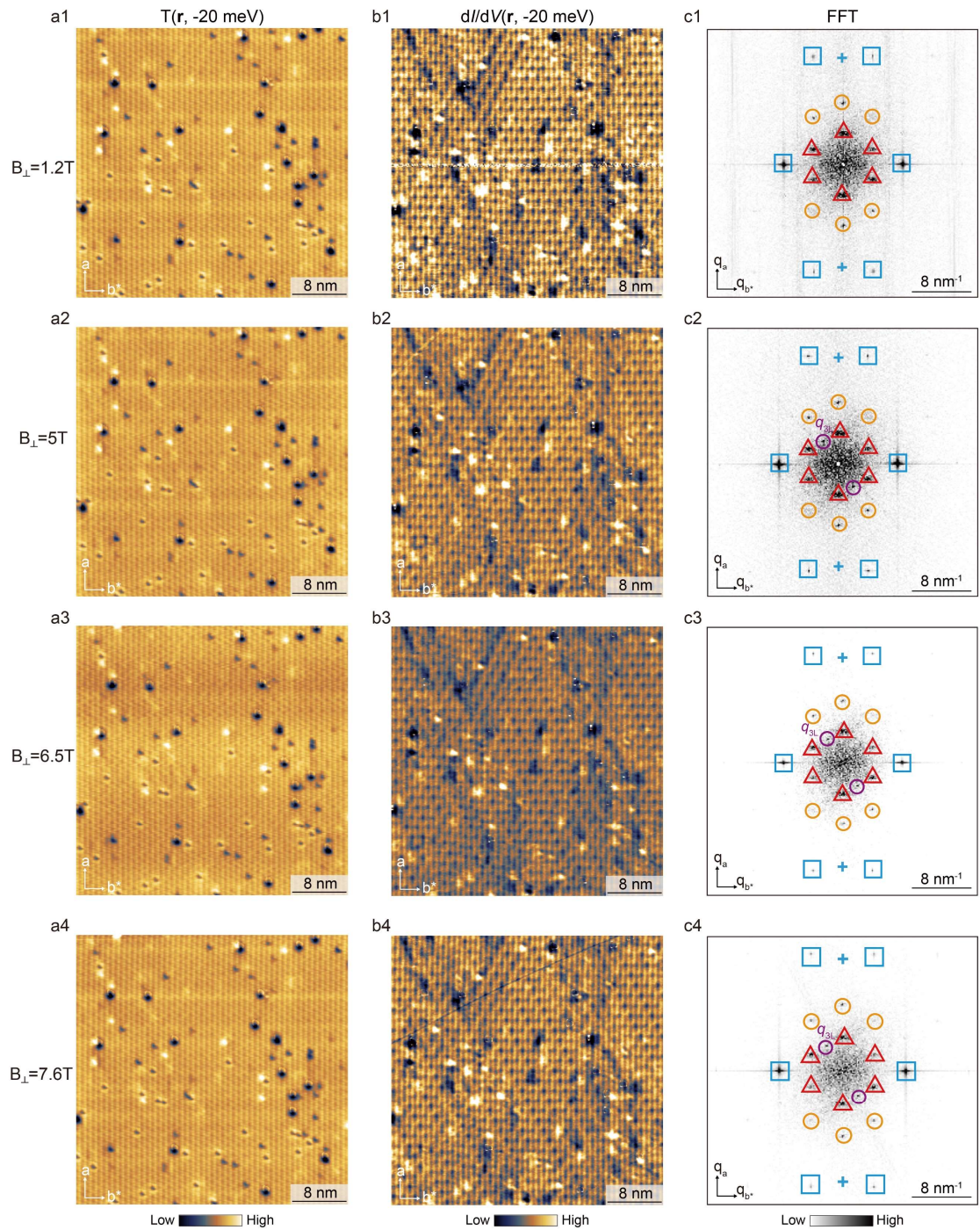

**Supplementary Figure 10 | Additional datasets for CO evolution with $B_\perp$ in sample #3, measured in the same FOV as Supplementary Fig. 8c.** (**a1**)-(**a4**), Representative topographic images acquired at various $B_\perp$. (**b1**)-(**b4**), Representative d$I$/d$V$ maps. (**c1**)-(**c4**), Corresponding FFT images of panels (**b1**)-(**b4**). Measurement conditions: (**a1**)-(**a4**), $V_b$ = -20 mV, $I_t$ = 200 pA; (**b1**)-(**b4**), $V_b$ = -20 mV, $I_t$ = 200 pA, $\Delta V$ = 5 mV.

**Supplementary Note 8: Spatial correlation between COs and magnetic vortices**

Supplementary Figs. 11a-d show zero-conductance maps at $\mathbf{B}_\perp$= 1.2 T, 3 T, 6.5 T, and 10 T, respectively, revealing the characteristic Yin-Yang vortex structure reported previously [5-7]. The Yang-parts of the vortices are highlighted by cyan rectangles in Supplementary Figs. 11a-c and indicated by arrows in Supplementary Fig. 11d. CO modulations measured in the same FOV under the same fields are presented in Supplementary Figs. 11e-h.

To examine the spatial correlation between COs and vortices, we extract the amplitude map for each CO component, $A(\mathbf{r}, |\mathbf{q}|)$, and overlay the positions of Yang-vortices (Supplementary Figs. 11i-u). No clear correlation between vortex locations and CO amplitude is observed. A similar lack of correlation is found for $\mathbf{q}_4$ wave vector (Supplementary Fig. 11v). Moreover, while COs vanish completely at 10 T (Supplementary Fig. 11h and Supplementary Figs. 3a1-a2), magnetic vortices remain clearly discernible (Supplementary Fig. 11d). Together, these observations indicate that the COs are largely decoupled from magnetic vortices.

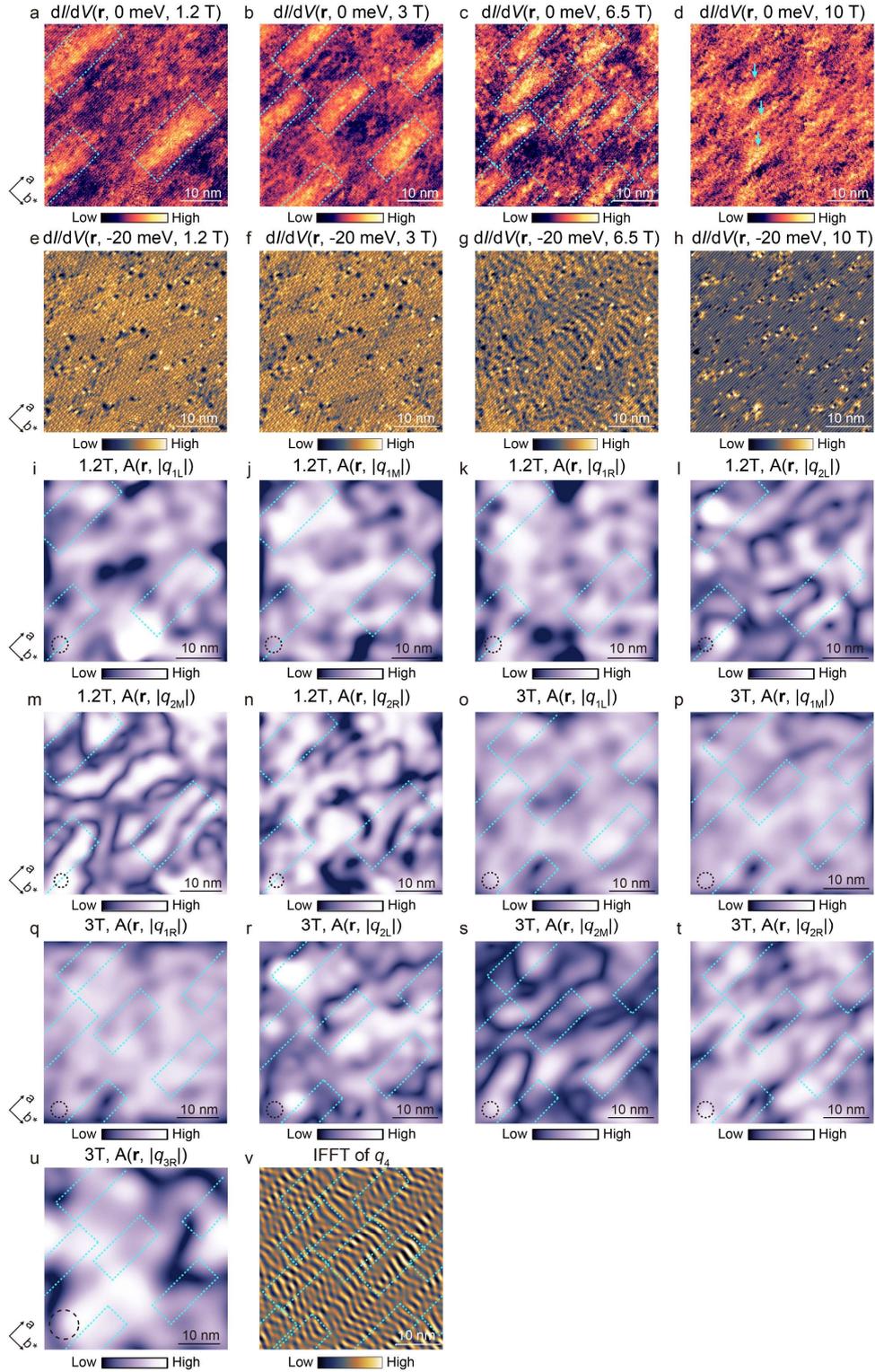

**Supplementary Figure 11 | Spatial correlation between COs and magnetic vortices.** (**a**)-(**d**), Vortex maps at $B_\perp$ = 1.2 T, 3 T, 6.5 T, and 10 T, respectively. Yang-part of vortices are indicated by cyan rectangles and arrows. (**e**)-(**h**), CO modulations measured in the same FOV under the same fields as panels (**a**)-(**d**). (**i**)-(**u**), Amplitude maps for different CO components. Dashed circles indicate the Gaussian FFT filter width ($\sigma_q$) used to obtain the amplitude maps. (**v**), Inverse fast Fourier transform (iFFT) image of $q_4$ component from panel (**g**). Measurement conditions: (**a**)-(**d**) $V_b$ = -1.5 mV, $I_t$ = 200 pA, $\Delta V$ = 0.08 mV; (**e**)-(**h**) $V_b$ = -20 mV, $I_t$ = 200 pA, $\Delta V$ = 5 mV.